\documentclass[11pt]{article}
\usepackage[a4paper,margin=2.6cm]{geometry}
\usepackage{amsmath,amssymb,amsthm}
\usepackage{mathtools}
\usepackage[hidelinks]{hyperref}
\usepackage{microtype}
\usepackage{graphicx}
\hypersetup{pdftitle={Neither graphs nor hypergraphs: structural moments of higher-order interactions determine stochastic social-impact dynamics},pdfauthor={Emanuele Cozzo}}

\newtheorem{proposition}{Proposition}
\newtheorem{theorem}[proposition]{Theorem}
\newtheorem{lemma}[proposition]{Lemma}
\newtheorem{corollary}[proposition]{Corollary}
\theoremstyle{definition}
\newtheorem{definition}{Definition}
\newtheorem{example}{Example}
\theoremstyle{remark}
\newtheorem{remark}{Remark}

\newcommand{\E}{\mathbb{E}}
\newcommand{\Qc}{\mathcal{Q}}
\newcommand{\Fc}{\mathcal{F}}
\newcommand{\Lg}{\mathcal{L}}
\newcommand{\Mo}[3]{M^{(#1,#2)}_{#3}}
\newcommand{\St}[2]{S(#1,#2)}

\title{Neither graphs nor hypergraphs: structural moments of higher-order interactions determine stochastic social-impact dynamics}
\author{Emanuele Cozzo$^{1,2}$\\[6pt]
\small $^{1}$Departament de F\'isica de la Mat\`eria Condensada, Universitat de Barcelona, Barcelona, Spain\\
\small $^{2}$Universitat de Barcelona Institute of Complex Systems (UBICS), Barcelona, Spain}
\date{\today}

\begin{document}
\maketitle

\begin{abstract}
Whether higher-order interactions can be reduced to simpler representations is a question about the dynamics they drive, not only about their structure: what can be discarded is whatever the dynamics does not see. We make this precise for binary-state stochastic dynamics in which each agent, when updated, interacts with a randomly selected group and changes state according to the fraction of disagreeing members, as in generalized voter and social-impact models on hypergraphs. We ask which statistics of the group-interaction process determine the resulting Markov dynamics. For response functions that are polynomials of degree $q$, the answer is a hierarchy of size-weighted inclusion moments, which we call structural moments, involving at most $q$ neighbors at a time, regardless of how large the groups are. These moments are not only sufficient but also necessary: two interaction structures produce identical dynamics for all such rules if and only if their moments coincide. More generally, each class of rules selects its own minimal structural description, ranging from weighted pairwise links for linear rules to the full distribution of co-membership by group size for arbitrary rules. For voter-type rules, the same moments also reconstruct exact pairwise copying representations of these dynamics. 
\end{abstract}

\section{Introduction}

Network science has long represented complex systems as collections of pairwise interactions. Many systems, however, are organized around interactions that involve several units at once, from scientific collaborations and face-to-face gatherings to neuronal assemblies and ecological communities. Hypergraphs and simplicial complexes have become the standard language for such group, or higher-order, interactions, and a large body of work has developed tools to describe their structure and to model the dynamics they support~\cite{battiston2020,battiston2021,bick}. The choice of representation is not neutral: graphs, hypergraphs and simplicial complexes encode different assumptions about how relations depend on one another, and these assumptions shape both the analysis and its interpretation~\cite{torres}.

Much of the interest in higher-order representations has come from dynamics. Models with group interactions display discontinuous transitions, bistability and critical-mass effects in social contagion~\cite{iacopini,dearruda}, abrupt desynchronization, multistability and explosive synchronization in coupled oscillators~\cite{skardal,millan}, departures from the average state in consensus dynamics~\cite{neuhauser2020}, and stabilization of large competitive communities in ecology~\cite{grilli}. These findings show that the way agents respond to their groups can reshape collective behavior. However, they do not show that interaction order is what produces it. Comparable phenomena arise in pairwise models with threshold, memory or cooperative transmission~\cite{doddswatts,cai} or with higher harmonics in the coupling~\cite{hansel,villegas}; whether a discontinuity survives fluctuations depends on dimensionality~\cite{meloni2026}; and in consensus dynamics, group interactions add nothing beyond rescaled pairwise ones unless the interaction function is nonlinear~\cite{neuhauser2020}. Moreover, the same node-level dynamics can often be written on a graph equipped with multivariate node functions~\cite{peixoto,rommens}. Collective phenomena therefore do not, by themselves, identify the interaction order or dictate a representation~\cite{villegas}. What they indicate is that the response of agents to their groups matters, which raises a sharper question: which features of the group-interaction process does a given dynamics actually depend on?

This question is at the center of a current debate on the reducibility of higher-order interactions. Peixoto et al.~\cite{peixoto} argue that graphs are maximally expressive, since any group dependence can be absorbed into multivariate functions on graph neighborhoods. Burgio~\cite{burgio} replies that expressive capacity is distinct from keeping interaction terms explicit, and that structure should be kept separate from the functional form of the interactions.

Other developments emphasize different notions of reducibility. Neuh\"auser et al.~\cite{neuhauser} study the effective order of hypergraph dynamics; Lucas et al.~\cite{lucas} consider dynamics-informed reducibility. Kirkley et al.~\cite{kirkley} quantify the structural reducibility of hypergraphs as the redundancy among their layers of different interaction sizes, and Xie et al.~\cite{xie} show that, under certain conditions, dynamics on higher-order networks can be transformed into dynamics on their pairwise projections. Arenas et al.~\cite{arenas} separate formal representability from adequacy for a given question, and argue that parsimony must be assessed for the combined structure--dynamics model, since information removed from the interaction structure can reappear as complexity in the dynamics. Villegas and Meloni~\cite{villegas} distinguish three levels at which interaction order can matter (reducibility of the coupling, the advantages of a higher-order representation, and identifiability from data) and argue that arity is not a universality label. Rommens et al.~\cite{rommens} prove that interaction order is a property of the dynamics rather than of its representation, formulating the choice between graphs and hypergraphs as a description-length problem. Lacasa~\cite{lacasa} examines the converse direction, showing that nonlinear dynamics on a graph can be read as linear dynamics on a higher-order structure obtained by lifting products of node states. P\'erez-Reche~\cite{perezreche} argues that reducibility is defined only relative to a target (a microscopic rule, a macroscopic observable, or a qualitative feature) and distinguishes representational, inferential, and mechanistic equivalence. These perspectives suggest that the information worth retaining depends on the dynamical rules and observables at issue.

We adopt a phenomenon-first stance. The primary object is the process of group interactions itself: which agents meet, in which groups, and how often. Graphs, hypergraphs, bipartite graphs and moment tensors are instruments for encoding this process, and the choice among them is subordinate to the question raised above. Accordingly, we describe interactions by a stochastic group-selection mechanism defined before any representation is chosen, and ask which of its statistics a dynamical rule depends on.

The exact mapping of hypergraph voter-type dynamics to state-dependent weighted pairwise copying processes by Llabr\'es et al.~\cite{llabres} offers a particularly transparent example. Here we consider a complementary reduction: rather than replacing group interactions with pairwise events, we seek state-independent statistics of the group-interaction process that determine the node-state Markov generator. We obtain a hierarchy of size-weighted moments, prove that it is sufficient for polynomial social-impact dynamics, and prove that it is also necessary, hence minimal, for natural classes of such dynamics. Whereas expressibility results establish what graph-based descriptions can represent, necessity identifies exactly what a class of dynamics requires. We connect the hierarchy to the effective copying weights, and extend the construction to arbitrary response functions, where it reveals that the bound on interaction order is specific to polynomial rules. Minimal sufficiency is an identifiability statement; it does not imply that the moments are the most compact encoding, a question we leave open.

\section{Structural moments of group interactions}\label{sec:moments}

Consider $N$ interacting agents. When agent $i$ is selected for an update, it interacts with a random group $G_i \subseteq V\setminus\{i\}$ drawn from a probability distribution $Q_i(G)$. We assume throughout that $Q_i$ does not depend on the dynamical states of the agents. It can represent the selection of an incident hyperedge, but it need not be specified through a hypergraph. We write $Q = \{Q_i\}_{i=1}^N$ for the group-selection mechanism, which describes the \emph{group-interaction process}: which groups each agent interacts with, and how often. We refer to it simply as the mechanism.

Introduce the random structural indicators $X_{ij} = \mathbf{1}\{j \in G_i\}$, their products $X_{iJ} = \prod_{j\in J} X_{ij}$, and the size of the selected group excluding $i$,
\begin{equation}
D_i = \sum_{j\neq i} X_{ij} .
\end{equation}
We assume that every selected group is nonempty, $D_i\ge 1$; on a hypergraph this requires every node to belong to at least one hyperedge with two or more nodes.
We denote by $D_{\max}$ the largest value of $D_i$ with positive probability, over all $i$.

The first- and second-order inclusion probabilities, $\E_Q[X_{ij}]$ and $\E_Q[X_{ij}X_{ik}]$, give the probability that an individual neighbor, or a pair of neighbors, participates in an interaction event with $i$. For dynamics that depend on the \emph{fraction} of agents in the selected group satisfying some condition, a more useful hierarchy includes the group size:
\begin{equation}
\Mo{r}{m}{i}(J) = \E_Q\!\left[\frac{X_{iJ}}{D_i^{m}}\right], \qquad J = \{j_1,\dots,j_r\},\quad 1 \le r \le m, \label{eq:moments}
\end{equation}
with $J$ a set of $r$ distinct indices different from $i$. We call these size-weighted inclusion moments \emph{structural moments}: they depend only on the mechanism $Q$, not on the dynamical state, and they capture correlations between group composition and group size.
The indices $r$ and $m$ have different roles: $r$ counts the distinct neighbors whose joint inclusion is recorded, while $m$ specifies the inverse power of group size. The weight $D_i^{-m}$ appears when a response to the fraction of group members meeting some condition is expanded in powers of that fraction: a term of degree $m$ contributes $D_i^{-m}$, and its expansion involves products of $r\le m$ distinct neighbors. Section~\ref{sec:voter} makes this connection explicit.
Since $X_{iJ} = 0$ whenever $|J| > D_i$, all moments with $r > D_{\max}$ vanish identically.

For an undirected hypergraph with incidence matrix $B_{ie}$ and node hyperdegree $\kappa_i$, if $G_i$ is obtained by selecting an incident hyperedge uniformly and excluding $i$, then
\begin{equation}
\Mo{r}{m}{i}(J) = \frac{1}{\kappa_i} \sum_{e \supseteq J \cup \{i\}} \frac{1}{(|e|-1)^m}, \label{eq:hypmoments}
\end{equation}
so that the moments are size-weighted co-membership statistics.

The role of the size weighting is governed by the number of distinct sizes in which a set $J$ appears together with $i$. If the group size seen by $i$ is deterministic, $D_i \equiv d_i$ (for instance, a node all of whose incident hyperedges have the same size), then $\Mo{r}{m}{i}(J) = d_i^{-m}\,\E_Q[X_{iJ}]$ and the size weighting is trivial: the hierarchy collapses onto plain inclusion probabilities. The full size information is contained in the \emph{size-resolved inclusion probabilities}
\begin{equation}
P_i(J,d) = \Pr_Q\!\left(J \subseteq G_i,\ D_i = d\right), \qquad d \ge |J|, \label{eq:sizeresolved}
\end{equation}
of which the moments are inverse power moments, $\Mo{r}{m}{i}(J) = \sum_d d^{-m} P_i(J,d)$. If the groups containing $J\cup\{i\}$ take $s$ distinct sizes and these sizes are known, the first $s$ moments of order $m \ge |J|$ determine $P_i(J,\cdot)$, since a distribution supported on $s$ known sizes is fixed by $s$ consecutive inverse power moments (a Vandermonde system in the inverse sizes $1/d$); higher moments carry no further information. If the sizes are not known, the moments with $m = |J|,\dots,D_{\max}$ determine $P_i(J,\cdot)$, since the same argument applies to all $D_{\max}-|J|+1$ candidate sizes. \footnote{This recovery is exact but numerically ill-conditioned when groups are large: as the interpolation points $1/d$ accumulate near zero, the columns of the Vandermonde matrix, with entries $d^{-m}$, become nearly parallel, so the matrix is close to singular and small errors in the moments are strongly amplified. Thus, exact equivalence of information does not imply general practical reconstructibility.}

Deterministic size is the case $s=1$. At the opposite extreme, each set $J$ that appears with $i$ does so in groups of every compatible size, $d = |J|,\dots,D_{\max}$, and moments up to order $D_{\max}$ are needed to recover $P_i(J,\cdot)$. The probabilities $P_i(J,d)$ are in fact a complete description of the mechanism: for any group $G$, a selected group of size $|G|$ that contains $G$ is $G$ itself, so that $P_i(G,|G|) = Q_i(G)$. Every truncation of the moment hierarchy is therefore a coarsening of $Q_i$. The point of the hierarchy is that particular dynamical classes require only a finite truncation. The truncation $\Pi_q$ is a genuine compression of $Q_i$ only when $q<D_{\max}$: for $q\ge D_{\max}$ it contains all moments with $m\le D_{\max}$ and therefore, by the argument above, determines $P_i(J,\cdot)$ for every $J$, and hence $Q_i$.

\section{Generalized voter dynamics}\label{sec:voter}

\subsection{Flip rates}

Consider binary-state agents, $s_i \in \{-1,+1\}$, and suppose that agent $i$ changes state with a probability determined by the fraction of disagreeing agents in its selected group. Let
\begin{equation}
a_{ij} = \frac{1 - s_i s_j}{2}
\end{equation}
be the disagreement indicator. Unlike $X_{ij}$, which describes the structure of an interaction event, $a_{ij}$ depends on the instantaneous dynamical configuration. The fraction of disagreeing agents in the selected group is $\phi_i = D_i^{-1} \sum_j X_{ij}a_{ij}$. For a social-impact function $f:[0,1]\to[0,1]$, the flipping probability conditional on selecting node $i$ is
\begin{equation}
r_i(s) = \E_Q\!\left[ f\!\left( \frac{\sum_j X_{ij} a_{ij}}{D_i} \right) \right]. \label{eq:rate}
\end{equation}
The requirement that $f$ map $[0,1]$ into $[0,1]$ ensures that $r_i$ is a probability. The expectation is taken over the group drawn at the update, conditional on the current configuration $s$. Since $Q_i$ does not depend on $s$, the disagreement indicators $a_{ij}$ are constants with respect to $\E_Q$ and factor out of it; the state-independence of group selection is the only independence assumption required. Equation~\eqref{eq:rate} separates the interaction structure, represented by the joint law of the $X_{ij}$, from the dynamical configuration, represented by the $a_{ij}$. This family includes the hypergraph voter models of Llabr\'es et al.~\cite{llabres}, as well as the $q$-voter-type rule $f(\phi)=\phi^q$~\cite{qvoter}.

We consider asynchronous updates in which node $i$ is selected at rate $p_i > 0$ and interacts with a group drawn independently from $Q_i$ at each update, so that the Markov generator $\Lg_{Q,f}$ has off-diagonal entries $p_i\, r_i(s)$ for the transition $s \to s^{(i)}$, where $s^{(i)}$ is $s$ with $s_i$ flipped.

For linear social impact, $f(\phi) = \phi$,
\begin{equation}
r_i(s) = \sum_j \Mo{1}{1}{i}(j)\, a_{ij}, \label{eq:linear}
\end{equation}
so the first size-weighted inclusion moments define the static weights of an equivalent voter model on a graph. For quadratic impact, $f(\phi) = \phi^2$, expanding Eq.~\eqref{eq:rate} and using $X_{ij}^2 = X_{ij}$ and $a_{ij}^2 = a_{ij}$ gives
\begin{equation}
r_i(s) = \sum_j \Mo{1}{2}{i}(j)\, a_{ij} + 2 \sum_{j<k} \Mo{2}{2}{i}(j,k)\, a_{ij} a_{ik}. \label{eq:quadratic}
\end{equation}

\subsection{Polynomial response}

Let $f(\phi) = \sum_{m=0}^{q} c_m \phi^m$. Since the summands $y_j = X_{ij}a_{ij}$ are binary, repeated factors collapse. For example,
\begin{equation*}
\Big(\sum_j y_j\Big)^3 = \sum_j y_j + 6\sum_{j<k}y_jy_k + 6\sum_{j<k<\ell} y_jy_ky_\ell.
\end{equation*}
Each coefficient counts the ways of assigning the three factors to a given set of one, two, or three distinct indices so that every index is used. At general order $m$, the same counting gives
\begin{equation}
\Big(\sum_j y_j\Big)^m = \sum_{r=1}^{m} r!\,\St{m}{r} \sum_{|J|=r} \prod_{j\in J} y_j , \label{eq:stirling}
\end{equation}
where $\St{m}{r}$ is a Stirling number of the second kind and $r!\,\St{m}{r}$ counts surjections from $m$ factors onto $r$ indices. Dividing by $D_i^m$ and averaging yields
\begin{equation}
r_i(s) = c_0 + \sum_{r=1}^{q} \sum_{\substack{J \subseteq V\setminus\{i\} \\ |J| = r}} K_i(J) \prod_{j \in J} a_{ij}, \label{eq:multilinear}
\end{equation}
with multilinear coefficients
\begin{equation}
K_i(J) = \sum_{m=|J|}^{q} c_m\, |J|!\, \St{m}{|J|}\, \Mo{|J|}{m}{i}(J). \label{eq:K}
\end{equation}
Each moment weights a group of size $d$ by $d^{-m}$, but the flip rate~\eqref{eq:multilinear} sums the coefficients over all sets $J$, and a group of size $d$ contains $\binom{d}{r}$ sets of $r$ members. The summed contribution of all its $r$-subsets to the degree-$m$ term can therefore scale as $d^{\,r-m}$, so large groups are not negligible in $r_i$.

We collect the moments up to order $q$ into the hierarchy
\begin{equation}
\Pi_q(Q) = \left\{ \Mo{r}{m}{i}(J) : 1 \le r \le m \le q \right\}. \label{eq:Piq}
\end{equation}

\begin{proposition}[Structural-moment sufficiency]\label{prop:sufficiency}
Let $Q$ and $Q'$ be state-independent group-selection mechanisms on the same node set, with the same node-selection rates $p_i > 0$. Let $f$ be a polynomial of degree at most $q$ with $f([0,1])\subseteq[0,1]$. If $\Pi_q(Q) = \Pi_q(Q')$, that is, $\Mo{r}{m}{i,Q}(J) = \Mo{r}{m}{i,Q'}(J)$ for every $i$, every $J$, and $1 \le r \le m \le q$, then $\Lg_{Q,f} = \Lg_{Q',f}$.
\end{proposition}

\begin{proof}
By Eqs.~\eqref{eq:multilinear}--\eqref{eq:K}, equality of the moments implies equality of every coefficient $K_i(J)$, hence of the flipping probabilities in every configuration. With common rates $p_i$, these determine the generator.
\end{proof}

The required moments have $r \le m \le q$. Their order $r$, the number of distinct neighbors involved, is therefore bounded by $\min(q, D_{\max})$: by $q$ because of the polynomial degree, and by $D_{\max}$ because moments with $r > D_{\max}$ vanish. The exponent $m$ is bounded by $q$ alone. A model involving arbitrarily large groups may therefore admit an exact description using low-order structural statistics. Conversely, the hierarchy is a genuine compression only when $q$ is small compared with the group sizes. If all groups selected by $i$ have the same size $d_i$ and $q \ge d_i$, then $\Pi_q$ contains the inclusion probabilities $\E_Q[X_{iJ}] = d_i^{m}\Mo{|J|}{m}{i}(J)$ for every $|J| \le d_i$, and in particular $Q_i(G) = \E_Q[X_{iG}]$ for every $|G| = d_i$: the moments then determine $Q_i$ completely. For a 3-uniform hypergraph, for instance, $\Pi_2$ already fixes the selection probabilities of the hyperedges incident to $i$. The reduction is informative in the regime $q < D_{\max}$.

\subsection{Arbitrary response functions}

The polynomial expansion is a special case of a general formula. For a fixed realized group $G$ of size $D$, the function $a \mapsto f\big(\sum_{j\in G} a_{j}/D\big)$ on $\{0,1\}^{G}$ has the multilinear expansion $\sum_{J \subseteq G} \prod_{j\in J}a_j\, \Delta_{1/D}^{|J|} f(0)$, where
\begin{equation}
\Delta_h^r f(0) = \sum_{k=0}^{r} (-1)^{r-k} \binom{r}{k} f(kh)
\end{equation}
is the $r$-th forward difference with step $h$.
For example, $\Delta_h^2 f(0)=f(2h)-2f(h)+f(0)$ measures the departure from additivity when two disagreeing neighbors each contribute a fraction $h$. More generally, the $r$-th difference isolates the joint contribution associated with $r$ distinct neighbors. Here $h=1/D$, so the joint contribution depends on group size; this is why the size-resolved probabilities appear in Proposition~\ref{prop:general} below.
The expansion itself follows from Newton's forward-difference formula, which is exact on the grid: with $n = \sum_{j\in G} a_j$ disagreeing members, $f(n/D) = \sum_{r=0}^{n} \binom{n}{r} \Delta^r_{1/D} f(0)$, and for binary variables $\binom{n}{r} = \sum_{J\subseteq G,\,|J|=r} \prod_{j\in J} a_j$, since both sides count the $r$-element subsets of disagreeing members. Equivalently, the coefficient of $J$ is the M\"obius inversion $\sum_{L\subseteq J}(-1)^{|J|-|L|} f(|L|/D)$ of the values of the rule on the Boolean cube $\{0,1\}^G$, which reduces to $\Delta^{|J|}_{1/D} f(0)$ because the rule depends only on the number of disagreeing members; the same inversion, applied to the full flip rate instead of a single group, recovers the coefficients $K_i(J)$ from the generator in Lemma~\ref{lem:mobius} below. Within a group, the coefficient of $J$ thus depends only on $|J|$, and any heterogeneity among the $K_i(J)$ originates in the structure, through which groups contain $J$ and their sizes. Averaging over $G_i$ gives the following.

\begin{proposition}[General multilinear coefficients]\label{prop:general}
For any $f:[0,1]\to[0,1]$, $r_i(s) = f(0) + \sum_{J\neq\emptyset} K_i(J) \prod_{j\in J} a_{ij}$ with
\begin{equation}
K_i(J) = \E_Q\!\left[ X_{iJ}\, \Delta_{1/D_i}^{|J|} f(0) \right] = \sum_{d \ge |J|} P_i(J,d)\, \Delta_{1/d}^{|J|} f(0). \label{eq:Kgeneral}
\end{equation}
\end{proposition}

For $f(\phi)=\phi^m$ one has $\Delta_h^r \phi^m(0) = h^m\, r!\,\St{m}{r}$, which recovers Eq.~\eqref{eq:K}. Since the rule is evaluated only at the accessible fractions $k/d$, what matters is not whether $f$ is globally polynomial but its restriction to these grids: on groups of size $d$ the expansion has degree $\max\{r\le d:\Delta^r_{1/d}f(0)\neq0\}$, which never exceeds $d$. For a generic non-polynomial response this degree equals $d$ for every accessible size. This is the case, for instance, for a threshold rule and for a nonlinear voter rule $\phi^\alpha$ with non-integer $\alpha$, whose $d$-th differences do not vanish, and Eq.~\eqref{eq:Kgeneral} then involves the size-resolved inclusion probabilities~\eqref{eq:sizeresolved} up to order $D_{\max}$. Conversely, a non-polynomial rule that coincides on all accessible grids with a polynomial of degree $q$ generates the same dynamics as that polynomial. The independence of the interaction order from group size is therefore a property of rules of bounded degree on the accessible grids, of which polynomial rules are the natural class, not of fraction-based dynamics in general.

\subsection{Operational interpretation}

In the original mechanism, an agent selects a group (for instance, one of its incident hyperedges), observes the opinions of its members, and updates according to the fraction of disagreeing agents. In the equivalent graph-based model, the agent evaluates the opinions of all its neighbors in the projected graph and combines them through the multilinear function~\eqref{eq:multilinear}, whose coefficients encode the statistical contribution of each neighbor, and of each combination of neighbors, across all possible interaction events. The graph-based model integrates out the randomness of group selection and directly evaluates the average flipping probability. The two processes have identical Markov generators, hence identical laws of the opinion configuration for identical initial distributions, although their microscopic interaction events need not coincide. The reduction does not eliminate higher-order dependencies; it transfers information about group composition into fixed coefficients of the effective rule, in line with the relocation of grouping information from structure to rule emphasized by Rommens et al.~\cite{rommens}.

\section{Identifiability and minimal sufficiency}

\subsection{Dynamical sufficiency}

\begin{definition}\label{def:sufficiency}
Let $\Qc$ be a class of state-independent group-selection mechanisms with common node-selection rates $p_i>0$, $\Fc$ a class of response functions, and $\Pi:\Qc\to\mathcal{R}$ a structural projection. We call $\Pi$ \emph{dynamically sufficient} for $\Fc$ if, for all $Q,Q'\in\Qc$,
\begin{equation}
\Pi(Q) = \Pi(Q') \;\Longrightarrow\; \Lg_{Q,f} = \Lg_{Q',f} \quad \forall f \in \Fc. \label{eq:sufficient}
\end{equation}
We call $\Pi$ \emph{minimal sufficient} if the converse implication also holds, i.e.\ $\Pi(Q)=\Pi(Q')$ if and only if the two mechanisms induce the same generator for every $f\in\Fc$. Equivalently, a minimal sufficient projection is a sufficient projection that is a function of every sufficient projection.
\end{definition}

The codomain $\mathcal{R}$ is arbitrary: only the equivalence relation $Q \sim_\Pi Q' \iff \Pi(Q) = \Pi(Q')$ induced by $\Pi$ matters, and sufficiency and minimality are properties of this relation. The terminology is borrowed from statistics, but here $\Pi$ is a function of the mechanism rather than of data: a minimal sufficient projection is the quotient of mechanisms by observational equivalence of the induced dynamics.

\begin{figure}[!htbp]
\centering
\includegraphics[width=\textwidth]{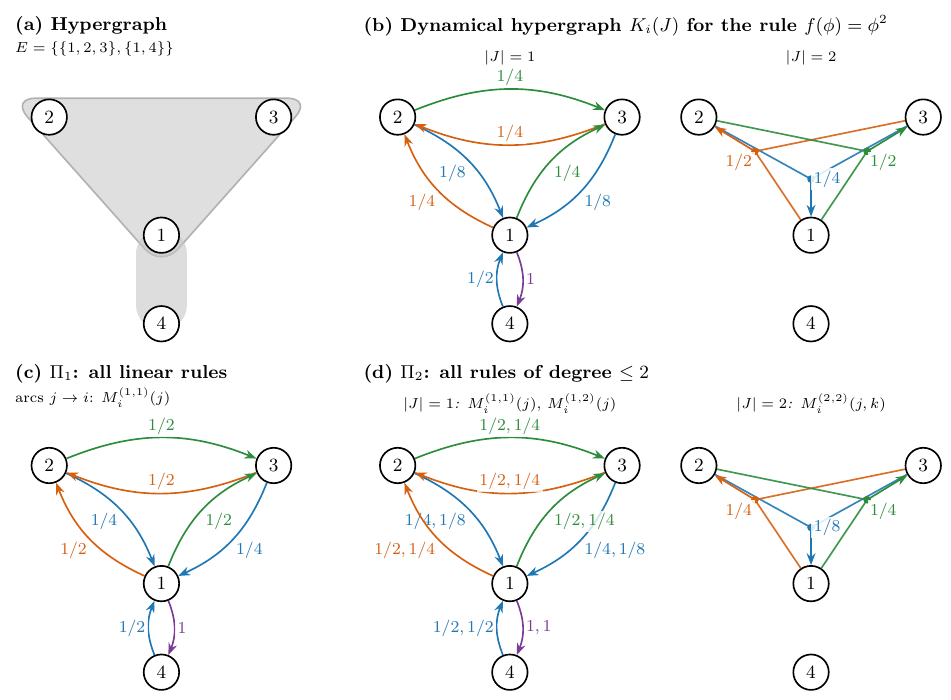}
\caption{Structural descriptions of a small hypergraph. (a) Hypergraph with hyperedges $\{1,2,3\}$ and $\{1,4\}$; when updated, each node selects one of its incident hyperedges uniformly at random. (b) Dynamical hypergraph of Remark~\ref{rem:canonical} for the rule $f(\phi)=\phi^2$: a hyperarc $J\to i$ carries the coefficient $K_i(J)$, here $K_i(j)=M^{(1,2)}_i(j)$ and $K_i(\{j,k\})=2M^{(2,2)}_i(j,k)$. Hyperarcs with $|J|=1$ and $|J|=2$ are drawn separately, and color identifies the head $i$; the coefficients entering each node sum to $f(1)=1$. (c) The hierarchy $\Pi_1$, a weighted directed graph with weights $M^{(1,1)}_i(j)$, is the minimal structural description for all linear rules. (d) The hierarchy $\Pi_2$ adds the size-weighted moments $M^{(1,2)}_i(j)$ and the co-membership moments $M^{(2,2)}_i(j,k)$, and is minimal for all rules of degree at most 2. All descriptions are directed although the hypergraph is not: for instance, $M^{(1,1)}_1(2)=1/4$ while $M^{(1,1)}_2(1)=1/2$, because node 1 belongs to two hyperedges and node 2 to one. Panels (b) and (d) can equivalently be read as labeled directed graphs, in which the edges of a hyperarc share its label and weight (Remark~\ref{rem:encodings}).}
\label{fig:example}
\end{figure}

The mechanisms $Q$ and $Q'$ may describe distinct hypergraphs or other stochastic group-selection processes; $Q'$ is not assumed to be a projection of $Q$. In this language, Proposition~\ref{prop:sufficiency} states that the hierarchy $\Pi_q$ of Eq.~\eqref{eq:Piq} is sufficient for polynomial rules of degree at most $q$. Figure~\ref{fig:example}(c,d) shows $\Pi_1$ and $\Pi_2$ for a small hypergraph. The following example illustrates both directions of Definition~\ref{def:sufficiency} for $q=1$.

\begin{example}[Three-body versus pairwise interactions]\label{ex:triangle}
Consider three nodes with equal selection rates and uniform selection of an incident hyperedge. Let $Q$ consist of the single hyperedge $\{1,2,3\}$, and $Q'$ of the three edges $\{1,2\}$, $\{1,3\}$, $\{2,3\}$. Under $Q$, node 1 always meets the group $\{2,3\}$, so $D_1 = 2$; under $Q'$, it meets $\{2\}$ or $\{3\}$ with probability $1/2$ each, and $D_1 = 1$. In both cases $\Mo{1}{1}{1}(2) = \Mo{1}{1}{1}(3) = 1/2$, and by symmetry $\Pi_1(Q) = \Pi_1(Q')$. For any linear rule $f(\phi) = c_0 + c_1\phi$,
\begin{equation}
r_1^{Q}(s) = c_0 + c_1\,\frac{a_{12}+a_{13}}{2} = \frac12\left(c_0 + c_1 a_{12}\right) + \frac12\left(c_0 + c_1 a_{13}\right) = r_1^{Q'}(s),
\end{equation}
and similarly for nodes 2 and 3. The two mechanisms generate the same dynamics for every linear rule: a three-body interaction is indistinguishable from its pairs. Conversely, adding the edge $\{1,2\}$ to $Q$ gives a mechanism $Q''$ with $\Mo{1}{1}{1}(2) = 3/4$ and $\Mo{1}{1}{1}(3) = 1/4$. If only node 2 disagrees with node 1 and $f(\phi) = \phi$, then $r_1^{Q} = 1/2$ while $r_1^{Q''} = 3/4$: different first-order moments are detected by a linear rule, as Theorem~\ref{thm:minimal} below asserts in general. Finally, the equivalence of $Q$ and $Q'$ does not survive nonlinear rules. For $f(\phi) = \phi^2$ and the same configuration, $r_1^{Q} = 1/4$ while $r_1^{Q'} = 1/2$, a difference carried by $\Mo{1}{2}{1}(2)$, which equals $1/4$ under $Q$ and $1/2$ under $Q'$.
\end{example}

For linear rules, the summaries $\Pi_1(Q)$ are weighted directed graphs with weights $\Mo{1}{1}{i}(j)$, which satisfy $\sum_j \Mo{1}{1}{i}(j) = 1$. Every equivalence class therefore contains a pairwise mechanism, in which $i$ selects a single neighbor $j$ with probability $\Mo{1}{1}{i}(j)$, in agreement with the exact static-weight mapping of the linear case in Ref.~\cite{llabres}.

In Example~\ref{ex:triangle}, the configuration considered separates the two mechanisms through the difference in their group sizes, and the distinguishing moment is $\Mo{1}{2}{i}$. (The mechanisms also differ in the co-membership moment $\Mo{2}{2}{1}(2,3)$, equal to $1/4$ under $Q$ and $0$ under $Q'$, which does not enter that configuration.) Second-order information also matters when all groups have the same size, through co-membership alone.

\begin{example}[Co-membership without size heterogeneity]\label{ex:octahedron}
Consider two 3-uniform hypergraphs on six nodes,
\begin{align}
E_A &= \{\{1,2,3\}, \{1,4,5\}, \{2,4,6\}, \{3,5,6\}\}, \\
E_B &= \{\{1,2,4\}, \{1,3,5\}, \{2,3,6\}, \{4,5,6\}\}.
\end{align}
In both hypergraphs every node belongs to two hyperedges, and every pair of nodes appears together in exactly one hyperedge, except the three pairs $\{1,6\}$, $\{2,5\}$, $\{3,4\}$, which never do. The two hypergraphs therefore have the same pairwise co-membership matrix $A_{ij} = \sum_e B_{ie}B_{je}$ ($i\ne j$), with unit weights on all other pairs. (Geometrically, the nodes are the vertices of an octahedron with opposite vertices $1$--$6$, $2$--$5$, $3$--$4$, and $E_A$ and $E_B$ are the two alternating sets of four faces.) Since all groups have the same size, every weighted projection of the form $\sum_e B_{ie}B_{je}\,h(|e|)$ coincides, as do all first-order moments $\Mo{1}{m}{i}(j)$, and linear voter dynamics are identical on the two hypergraphs. The co-membership moments differ: node 1 belongs to the groups $\{2,3\}$, $\{4,5\}$ in $A$ and to $\{2,4\}$, $\{3,5\}$ in $B$, so that $\Mo{2}{2}{1}(2,3) = 1/8$ in $A$ and $0$ in $B$. If nodes 2 and 3 disagree with node 1 while the remaining nodes agree with it, the quadratic rule $f(\phi)=\phi^2$ gives
\begin{equation}
r_1^{A} = \frac12\left(1^2 + 0^2\right) = \frac12, \qquad
r_1^{B} = \frac12\left[\left(\frac12\right)^2 + \left(\frac12\right)^2\right] = \frac14. \label{eq:counter}
\end{equation}
\end{example}

The two examples display the two channels through which second-order structural information enters nonlinear dynamics, group-size heterogeneity and co-membership. In both, mechanisms with the same pairwise co-membership counts generate distinct nonlinear dynamics. Theorem~\ref{thm:minimal} below shows that this holds in general: whenever two mechanisms share $\Pi_1$ but differ in some component of $\Pi_q$, whether a size-weighted first-order moment $\Mo{1}{m}{i}$ with $2\le m\le q$, as in Example~\ref{ex:triangle}, or a co-membership moment $\Mo{r}{m}{i}$ with $r\ge 2$, as in Example~\ref{ex:octahedron}, some polynomial rule of degree at most $q$ satisfying the voter endpoint conditions distinguishes them (Corollary~\ref{cor:classes}). That distinct hypergraphs can share their co-occurrence matrix is known~\cite{larock}; the pair in Example~\ref{ex:octahedron} is the Pasch trade of design theory, the smallest pair of triple systems covering the same pairs~\cite{colbourn}. What the example adds is that equal weighted co-membership, together with equal size-weighted first-order moments, does not fix nonlinear dynamics even under a fraction-based rule, which aggregates disagreement additively within the selected group. A deterministic analogue is given in Ref.~\cite{rommens}, where a single 4-hyperedge and the four triangles on its nodes share the complete graph as projection but generate different unanimity-type activation dynamics; there the two structures differ in how nodes are grouped, a difference made visible by a rule that aggregates disjunctively over groups; here they are separated by co-membership moments under additive aggregation.

We now show that $\Pi_q$ is also necessary.

\subsection{Uniqueness of the multilinear coefficients}

\begin{lemma}\label{lem:mobius}
The generator $\Lg_{Q,f}$ determines every coefficient $K_i(J)$ in Eq.~\eqref{eq:multilinear} (or in Proposition~\ref{prop:general} for general $f$), through
\begin{equation}
K_i(J) = \sum_{L \subseteq J} (-1)^{|J|-|L|}\, r_i\big(s^{[i,L]}\big), \qquad J \neq \emptyset, \label{eq:mobius}
\end{equation}
where $s^{[i,L]}$ is any configuration in which the agents disagreeing with $i$ are exactly those in $L$.
\end{lemma}

\begin{proof}
The generator determines $r_i(s) = \Lg_{Q,f}(s,s^{(i)})/p_i$ for every $s$. Fix $s_i$. The map $s \mapsto (a_{ij})_{j\neq i}$ is then a bijection onto $\{0,1\}^{N-1}$, so $r_i$ is a function on the hypercube, and every such function has a unique multilinear representation~\cite{odonnell}. Evaluating Eq.~\eqref{eq:multilinear} at $a = \mathbf{1}_L$ gives $r_i(s^{[i,L]}) = c_0 + \sum_{\emptyset\neq J\subseteq L} K_i(J)$, and M\"obius inversion on the Boolean lattice yields Eq.~\eqref{eq:mobius}; the constant term cancels because $\sum_{L\subseteq J}(-1)^{|J|-|L|} = 0$ for $J\neq\emptyset$.
\end{proof}

Equation~\eqref{eq:mobius} is also an operational recipe: $K_i(J)$ is the inclusion--exclusion combination of the flip rates of $i$ over the $2^{|J|}$ configurations in which only members of $J$ disagree with it. The lemma is the standard uniqueness of the Fourier--Walsh, or multilinear, expansion of a function on the Boolean cube~\cite{odonnell}. Rommens et al.~\cite{rommens} use the same expansion and inversion formula for deterministic dynamical maps to show that the sets of variables acting jointly in a node's update, and hence its interaction order, are fixed by the map and identical in every exact representation. Here the expansion serves a further purpose: the coefficients $K_i(J)$ are combinations of structural moments, Eq.~\eqref{eq:K}, so their identifiability from the generator is the first step towards identifying which statistics of the group-interaction process the dynamics determines (Theorem~\ref{thm:minimal}).

\begin{corollary}[Single rule]\label{cor:single}
For a fixed response $f$, the projection $\Pi_f(Q) = \{K_i(J)\}_{i,J}$ is minimal sufficient for $\Fc = \{f\}$.
\end{corollary}

Thus, for a single rule, the question of which combinations of moments matter has an exact answer: exactly the combinations~\eqref{eq:K}, or~\eqref{eq:Kgeneral} for general $f$, and no coarser statistic.

\subsection{Necessity of the moment hierarchy}

For a class $\Fc$ of polynomials of degree at most $q$, let
\begin{equation}
V(\Fc) = \operatorname{span}\left\{ (c_1,\dots,c_q) : f = \textstyle\sum_{m=0}^q c_m \phi^m \in \Fc \right\} \subseteq \mathbb{R}^q .
\end{equation}

Sufficiency follows from the expansion of the flip rates, but necessity requires separating moments that enter the same multilinear coefficient. For $q=2$, for instance,
\begin{equation*}
K_i(\{j\})=c_1\Mo{1}{1}{i}(j)+c_2\Mo{1}{2}{i}(j).
\end{equation*}
For one fixed rule, equality of this coefficient constrains only one combination of two moments. Requiring equality for different response functions supplies additional linear constraints. If their coefficient vectors span the polynomial coefficient space, those constraints force each moment to agree separately. Theorem~\ref{thm:minimal} makes this argument for arbitrary $q$.

\begin{theorem}[Minimal sufficiency of $\Pi_q$]\label{thm:minimal}
If $V(\Fc) = \mathbb{R}^q$, then $\Pi_q$ is minimal sufficient for $\Fc$.
\end{theorem}

\begin{proof}
Sufficiency is Proposition~\ref{prop:sufficiency}. For necessity, suppose $\Lg_{Q,f} = \Lg_{Q',f}$ for all $f\in\Fc$. By Lemma~\ref{lem:mobius}, $K_{i,Q}(J) = K_{i,Q'}(J)$ for every $f\in\Fc$, $i$ and $J$. Fix $i$ and $J$ with $|J| = r \le q$, and set $g_m = r!\,\St{m}{r}\big[\Mo{r}{m}{i,Q}(J) - \Mo{r}{m}{i,Q'}(J)\big]$ for $m \ge r$ and $g_m = 0$ for $m < r$. By Eq.~\eqref{eq:K}, the linear functional $c \mapsto \sum_{m=1}^q c_m g_m$ vanishes on every coefficient vector of $\Fc$, hence on their span $\mathbb{R}^q$, so $g = 0$. Since $\St{m}{r} > 0$ for $m \ge r$, the moments coincide.
\end{proof}

The theorem applies to the natural classes of social-impact rules.

\begin{corollary}\label{cor:classes}
Let $q \ge 1$. $\Pi_q$ is minimal sufficient for
\begin{enumerate}
\item[(a)] $\Fc_q$: all polynomials of degree at most $q$ with $f([0,1]) \subseteq [0,1]$;
\item[(b)] $\Fc_q^{\mathrm{v}}$: the polynomials in $\Fc_q$ satisfying the voter endpoint conditions $f(0) = 0$, $f(1) = 1$.
\end{enumerate}
\end{corollary}

\begin{proof}
(a) The polynomials $\tfrac12 + \tfrac12\phi^m$ belong to $\Fc_q$ for $m=1,\dots,q$. Their coefficient vectors are $\tfrac12 e_m$, which span $\mathbb{R}^q$.

(b) \emph{Admissibility.} The class $\Fc_q^{\mathrm{v}}$ contains $\phi$ and, for $k=0,\dots,q-2$ and $0<\varepsilon\le 1$, the functions $f_k=\phi+\varepsilon\,\phi^{k+1}(1-\phi)$. They satisfy the endpoint conditions; moreover, $f_k=\phi[1+\varepsilon(1-\phi)\phi^k]\ge 0$ and $1-f_k=(1-\phi)(1-\varepsilon\phi^{k+1})\ge 0$ on $[0,1]$.

\emph{Spanning.} Their coefficient vectors are $e_1+\varepsilon(e_{k+1}-e_{k+2})$, while the vector for $\phi$ is $e_1$. Together they span the same space as $e_1,e_1-e_2,\dots,e_{q-1}-e_q$, namely $\mathbb{R}^q$.
\end{proof}

Note that the endpoint conditions confine the coefficients to an affine hyperplane $\sum_m c_m = 1$ not passing through the origin. Because generator equality is a \emph{linear} condition on the coefficients, vanishing on this hyperplane forces vanishing on its linear span, which is all of $\mathbb{R}^q$; no residual non-identifiability survives.

The same argument applies beyond polynomials. For the class of all response functions $f:[0,1]\to[0,1]$, the functionals $f \mapsto \Delta_{1/d}^{r} f(0)$ are linearly independent across $d$: the point $1/d_{\max}$ enters only the functional with the largest $d$, with coefficient $(-1)^{r-1} r \ne 0$, and induction on $d$ completes the argument. Hence the size-resolved inclusion probabilities~\eqref{eq:sizeresolved}, and with them the mechanism $Q$ itself, are minimal sufficient for unrestricted fraction-based dynamics: when every response function is admissible, the dynamics determines the group-interaction process completely and no structural reduction is possible. The moment hierarchy truncated at order $q$ is the exact reduction induced by restricting to degree-$q$ polynomials.

\subsection{Dynamical hypergraph and equivalent encodings}\label{sec:encodings}

The multilinear coefficients identified in Lemma~\ref{lem:mobius} admit a direct structural reading, which connects the results above with representational questions.

\begin{remark}[Dynamical hypergraph]\label{rem:canonical}
The flip rate~\eqref{eq:multilinear} is linear in the product variables $\prod_{j\in J} a_{ij}$: in the spirit of Lacasa's lifting~\cite{lacasa}, it carries one weight $K_i(J)$ on each hyperedge $J\cup\{i\}$ of a weighted hypergraph. Unlike a Carleman linearization, this statement concerns the rate, not a closed linear evolution of the lifted variables, which would require all products of spins. Since the weights depend on the focal node, this hypergraph is directed, with one hyperarc $J\to i$ per interaction term [Fig.~\ref{fig:example}(b)]. Directed hypergraphs of this kind, with the focal node as the head of each set, were proposed by Burgio~\cite{burgio} as a representation of interaction terms, in close relation to the focal hypergraphs of Moutuou~\cite{moutuou}, and arise in Rommens et al.~\cite{rommens} as the incidence structure induced by a decomposed update rule; Rommens et al.\ also showed that, for deterministic binary maps, the sets carrying nonzero coefficients are fixed by the map alone. Here they are fixed by the generator. Two features are specific to random group selection. First, the hypergraph is directed even when $Q$ is an undirected hypergraph: under uniform hyperedge selection the moments~\eqref{eq:hypmoments} carry a factor $1/\kappa_i$, so hyperarc weights depend on the hyperdegree of their head. Second, every hyperarc lies within a group: by Eq.~\eqref{eq:Kgeneral}, $K_i(J)=0$ unless $J\subseteq G_i$ with positive probability, because the flip rate averages over a single selected group. Non-additive aggregation over groups, such as a disjunction, can instead produce interaction terms contained in no group~\cite{rommens}. For general polynomial nonlinearities, such liftings require hb-graphs~\cite{lacasa}, i.e.\ hyperedges with repeated vertices, because monomials such as $x_j^2 x_k$ appear. For binary variables, idempotence $a_{ij}^2 = a_{ij}$ collapses every multiset onto its support, and the factor $|J|!\,\St{m}{|J|}$ in Eq.~\eqref{eq:K} counts the ordered $m$-tuples (surjections) with support $J$, that is, the multisets of size $m$ with support $J$ weighted by their multinomial coefficients.
\end{remark}

\begin{remark}[Equivalent encodings]\label{rem:encodings}
Lemma~\ref{lem:mobius} implies that, once the disagreement indicators of each focal node are chosen as variables, the dynamical hypergraph of Remark~\ref{rem:canonical} is uniquely determined by the generator. These variables expand the rate around the configuration in which all agents agree with $i$, in the same way as interaction terms are defined relative to a reference value of each variable in Ref.~\cite{burgio}. The choice matters only for the lower-order terms: a change of variables, for instance to the spins $s_j$, transforms the coefficients by an invertible linear map that preserves the interaction order and the inclusion-maximal hyperarcs but can change the lower-order ones, in line with the corresponding statement for relabeled binary states in Ref.~\cite{rommens}. Decompositions into non-multilinear interaction functions, or liftings to enlarged state spaces as in Ref.~\cite{lacasa}, are not unique; such liftings establish representational rather than explanatory equivalence~\cite{villegas}. Even within a fixed basis, the directed hypergraph is one of several equivalent encodings. Each hyperarc $J\to i$ can be replaced by the $|J|$ directed edges $j\to i$, $j\in J$, all labeled by $J$ and carrying the weight $K_i(J)$. The resulting labeled, weighted, directed multigraph has as its skeleton the directed graph of pairs $j\to i$ with $j$ in some $J$ such that $K_i(J)\neq 0$, generically the pairwise projection, and carries the same information, provided that edges sharing a label are constrained to share their head and their weight: the directed hypergraph builds this constraint into the representation, whereas the labeled graph imposes it on the labels. The labels index sets of neighbors, as the coefficients $K_i(J)$ require, and not the groups in which the interactions originate. A related comparison, between an edge-colored clique expansion and a factor graph, is made in Ref.~\cite{rommens}. What is invariant across bases and encodings is the induced equivalence: two mechanisms have the same coefficients in one basis if and only if they do in any other, so the equivalence classes of Theorem~\ref{thm:minimal} depend neither on the choice of variables nor on the encoding.
\end{remark}

\subsection{Scope}

The equivalence is exact at the level of the Markov generator. Equality of generators implies equality of all finite-dimensional distributions of the node-state process, hence of all equal- and unequal-time correlations, for identical initial distributions. This holds irrespective of whether the hierarchy of correlation equations closes. For $q=1$, the generator maps products of $k$ spins into products of at most $k$ spins, so correlations of bounded order obey closed equations, which is the origin of the exact solvability of the linear voter model. When the rule has effective multilinear degree larger than one on the groups that occur, the generator maps them into products of up to $k+q$ spins and the hierarchy in general does not close; a polynomial rule of nominal degree $q\ge 2$ that is linear on the accessible disagreement fractions, such as the cubic rule discussed in Sec.~\ref{sec:discussion} on groups with $D=2$, still yields a closed hierarchy. Non-closure concerns the tractability of a given process; two equivalent mechanisms share the same open hierarchy with the same coefficients.

The equivalence applies to the exact dynamics, not to approximations of it. Closure schemes derived from the generator, such as mean-field or pair approximations of the moment equations, inherit the equivalence. Schemes built directly from the interaction structure, such as hyperedge-based approximate master equations, may use information beyond $\Pi_q$ and can then assign different predictions to equivalent mechanisms. A closure is consistent with the exact equivalence classes only if it depends on the structure through $\Pi_q$.

The equivalence does not preserve the statistics of interaction events themselves: if group membership, joint exposure, or participation in specific events are observables of interest, a richer structural description is required. Minimal sufficiency is also distinct from parsimony. Higher-order moment tensors may be dense, and the full interaction structure may provide a more compact parametrization; Theorem~\ref{thm:minimal} identifies \emph{which} information is dynamically relevant, not the shortest way to store it.

The mechanism $Q=\{Q_i\}$ specifies, for each focal agent, the distribution of the group it interacts with at an update, not correlations between group-selection events. The results above apply whenever these marginals determine the transition rates: in the asynchronous dynamics considered here, and in synchronous updating with independent group selections and flip decisions, where joint transition probabilities factorize into single-agent rates. Correlated selection may require a richer structural description. If several agents update in a common event, their joint transition probabilities depend on cross-agent moments $\E[X_{iJ}X_{kL}/(D_i^mD_k^n)]$, which are not in general determined by the hierarchies of the individual marginals. Temporal correlations between successive selections, such as a group that persists over several updates, can make the node-state process non-Markovian, and a Markov description then requires the current groups as part of the state. The structural information a dynamics requires thus depends not only on the response function but also on which correlations of the interaction process the dynamics can access.

\section{Connection with exact pairwise reductions}

The preceding sections identify the structural moments that are sufficient and necessary to determine the multivariate flip rates. We now show that the same moments also determine an exact pairwise copying representation. This connects the hierarchy to the reduction of Llabr\'es et al.~\cite{llabres} while keeping distinct a pairwise realization of the dynamics and the state-independent structural information that determines it.

In that construction, the generalized voter dynamics is represented through pairwise copying probabilities $w_{ij}(s)$ satisfying
\begin{equation}
r_i(s) = \sum_j w_{ij}(s)\, a_{ij}, \qquad \sum_j w_{ij}(s) = 1. \label{eq:copying}
\end{equation}
Their construction first samples a group and then assigns equal copying probabilities to all members in each opinion class. In our notation,
\begin{equation}
w_{ij}(s) = \E_Q\!\left[ \frac{X_{ij}}{D_i} \left( a_{ij}\, \frac{f(\phi_i)}{\phi_i} + (1-a_{ij})\, \frac{1-f(\phi_i)}{1-\phi_i} \right) \right], \label{eq:llabres}
\end{equation}
where, under the endpoint conditions $f(0)=0$, $f(1)=1$, the ratios are understood as the polynomials obtained by division, so that the integrand is well defined at $\phi_i \in \{0,1\}$. For every realized group, summing the integrand over $j$ gives $\phi_i f/\phi_i + (1-\phi_i)(1-f)/(1-\phi_i) = 1$, and summing it against $a_{ij}$ gives $f(\phi_i)$, which establishes Eq.~\eqref{eq:copying}.

For $f(\phi) = \phi$ the weights are static, $w_{ij} = \Mo{1}{1}{i}(j)$. For $f(\phi) = \phi^2$, the integrand becomes $X_{ij}(1 - a_{ij} + \phi_i)/D_i$. Expanding $\phi_i$, and separating the term $k=j$, which contributes $X_{ij}^2 = X_{ij}$, gives
\begin{equation}
w_{ij}(s) = (1 - a_{ij})\, \Mo{1}{1}{i}(j) + \Mo{1}{2}{i}(j)\, a_{ij} + \sum_{k \neq j} \Mo{2}{2}{i}(j,k)\, a_{ik}. \label{eq:wquad}
\end{equation}
Multiplying by $a_{ij}$ and summing over $j$ recovers Eq.~\eqref{eq:quadratic}, since $(1-a_{ij})a_{ij} = 0$.

\paragraph{General polynomial response.}
Let $f$ have degree $q$ and satisfy the endpoint conditions, so $c_0 = 0$ and $\sum_m c_m = 1$. Then
\begin{equation}
\frac{f(\phi)}{\phi} = \sum_{m=1}^{q} c_m \phi^{m-1}, \qquad
\frac{1-f(\phi)}{1-\phi} = \sum_{n=0}^{q-1} b_n \phi^n, \quad b_n = 1 - \sum_{m=1}^{n} c_m ,
\end{equation}
where the second identity follows from multiplying $1 - f$ by the geometric series and using $\sum_m c_m = 1$ to truncate it. Both ratios are therefore polynomials in $\phi_i$, and the integrand of Eq.~\eqref{eq:llabres} is a combination of the terms $X_{ij}\phi_i^{\,n}/D_i$. Each of these is handled exactly as the flip rate in Sec.~\ref{sec:voter}: writing $\phi_i^{\,n} = D_i^{-n}\big(\sum_k X_{ik}a_{ik}\big)^n$ and expanding with Eq.~\eqref{eq:stirling} produces products of disagreement indicators over sets $L$ of neighbors, and averaging over the selected group turns their structural prefactors into size-weighted moments. The only differences are that the factor $X_{ij}$ adds $j$ to each set and the factor $1/D_i$ raises the exponent by one. This yields the state-dependent structural sums
\begin{equation}
T^{(n)}_{ij}(s) = \E_Q\!\left[ \frac{X_{ij}\, \phi_i^{\,n}}{D_i} \right] = \sum_{\ell=0}^{n} \ell!\, \St{n}{\ell} \sum_{\substack{L \subseteq V\setminus\{i\} \\ |L| = \ell}} \Mo{|L\cup\{j\}|}{n+1}{i}\!\big(L \cup \{j\}\big) \prod_{k\in L} a_{ik}, \label{eq:T}
\end{equation}
with the convention $0!\,\St{0}{0}=1$. Then
\begin{equation}
w_{ij}(s) = a_{ij} \sum_{m=1}^{q} c_m\, T^{(m-1)}_{ij}(s) + (1-a_{ij}) \sum_{n=0}^{q-1} b_n\, T^{(n)}_{ij}(s). \label{eq:wgeneral}
\end{equation}
Equation~\eqref{eq:wquad} is the case $c_1=0$, $c_2=1$, for which $b_0=b_1=1$. The moments entering Eq.~\eqref{eq:T} have order at most $n+1 \le q$, so $\Pi_q$ suffices both to specify the multivariate flip rule and to reconstruct this particular exact pairwise realization. The copying representation is not unique, and its configuration-dependent weights must not be confused with purely structural edge weights.

\section{Discussion}\label{sec:discussion}

The structural-moment construction is a concrete realization of dynamics-dependent structural reducibility. The results organize into a hierarchy of minimal sufficient statistics indexed by the dynamical class. For a single rule, the minimal statistic is the set of multilinear coefficients $K_i(J)$, recoverable from the generator by M\"obius inversion. For polynomial rules of degree at most $q$, with or without the voter endpoint conditions, it is the moment hierarchy $\Pi_q$, of order at most $\min(q, D_{\max})$. For arbitrary fraction-based rules, it is the set of size-resolved inclusion probabilities, which is equivalent to the mechanism itself. Coarser rule classes thus induce coarser equivalence classes of interaction structures, and in each case the reduction is exact and cannot be improved without losing dynamical information. In the taxonomy of P\'erez-Reche~\cite{perezreche}, this is a representational equivalence with respect to a mechanistic target in his sense, the generator itself, with zero tolerance, and it is settled mathematically rather than inferentially. It also bounds what can be inferred from node-state trajectories. If the rule and the rates $p_i$ are known, the trajectories determine at most the coefficients $K_i(J)$, which are fixed combinations of the moments; recovering $\Pi_q$ itself requires observing the same mechanism under rules whose coefficient vectors span $\mathbb{R}^q$. If the rule is unknown, structure and rule are partly confounded and less is identifiable. In every case, two mechanisms with equal $\Pi_q$ are indistinguishable under all rules in $\Fc_q$, regardless of the amount of data. Theorem~\ref{thm:minimal} thus characterizes in advance, for this class of dynamics, the equivalence classes of mechanisms that generate the same observables, as called for in Ref.~\cite{perezreche}.

\paragraph{Expressibility and explicit structure.}
The construction is consistent with the expressibility result of Peixoto et al.~\cite{peixoto}: every dynamics considered here is exactly a multivariate function on the projected graph. Examples~\ref{ex:triangle} and~\ref{ex:octahedron} are therefore not counterexamples to that result; they show instead that a graph's structural weights, together with a prescribed social-impact function, do not in general determine those multivariate functions. The equivalence noted in Remark~\ref{rem:encodings} makes this relation explicit: the multivariate functions are encoded exactly by labels on the edges of the projected graph, a precise version of the colored links invoked in Ref.~\cite{peixoto}. Whether these labels are written on a graph or organized into hyperarcs is a choice of instrument; what the dynamics fixes is the equivalence class, not its encoding. Burgio~\cite{burgio} argues that structure should be kept separate from the functional form of interactions and proposes directed hypergraphs, with one hyperarc per interaction term; the same structure arises in Ref.~\cite{rommens} as the incidence structure induced by a decomposed update rule. The coefficients $K_i(J)$ realize this representation, with hyperarcs $J\to i$, and Eq.~\eqref{eq:K} makes the separation explicit: each coefficient combines rule-dependent constants with purely structural moments. The hierarchy $\Pi_q$ is the part of the structural side that a class of rules requires. Remark~\ref{rem:canonical} connects both views with Lacasa's observation~\cite{lacasa} that nonlinear graph dynamics can be read as linear dynamics on a higher-order structure. For binary states, the lifting is finite and involves ordinary hypergraphs rather than hb-graphs; once the disagreement indicators are chosen as variables, it is also unique. Within this representation the dynamics singles out a weighted hypergraph, and an interaction structure matters only through its image under the map $Q \mapsto \{K_i(J)\}$, or $Q \mapsto \Pi_q$ for a class of rules. This identifies what the dynamics determines, not a preferred representation among all possible ones, consistent with the representational freedom emphasized in Ref.~\cite{lacasa}. The dichotomy between finite and infinite liftings in Ref.~\cite{lacasa} reappears here in the correlation hierarchy, which closes when the flip rate is linear in the disagreement indicators and in general does not otherwise.

\paragraph{Exact pairwise reductions.}
Llabr\'es et al.~\cite{llabres} showed that generalized voter dynamics on hypergraphs admit exact representations as pairwise copying processes with configuration-dependent weights, and, within a pair approximation supported by simulations on random hypergraphs, that the macroscopic dynamics of the linear case coincides with that of the standard voter model on the unweighted projected network. The present results complement theirs in three respects. We characterize the state-independent structural data that determine the dynamics, rather than a particular realization of it. We prove that these data are necessary as well as sufficient. And we show that the same data reconstruct their copying weights for any polynomial rule, Eq.~\eqref{eq:wgeneral}. The state-dependent weights $w_{ij}(s)$ and the multivariate rule~\eqref{eq:multilinear} are two exact realizations of one equivalence class.

\paragraph{Effective order.}
Neuh\"auser et al.~\cite{neuhauser} define the dynamical order of a hypergraph dynamical system as the smallest $p$ such that every group interaction decomposes into a sum of terms involving the focal node and at most $p-1$ other members. For binary states, a symmetric function of the members of a group admits such a decomposition if and only if its multilinear degree is at most $p-1$~\cite{rommens}. By Proposition~\ref{prop:general}, the multilinear degree of the rules studied here on a group with $D$ other members is the largest $r\le D$ such that $\Delta^{r}_{1/D}f(0)\neq 0$, so their dynamical order is at most $\min(q,D)+1$, in agreement with the bound stated after Proposition~\ref{prop:sufficiency}. Equality holds whenever $D\ge q$, since $\Delta^q_h f(0)=q!\,c_q h^q\neq 0$, and generically otherwise; it fails when $f$ coincides on the grid $\{0,1/D,\dots,1\}$ with a polynomial of lower degree. For instance, $f(\phi)=\phi+\phi(\phi-\tfrac12)(\phi-1)$ acts as the linear voter rule on groups with $D=2$. Applied to the present rules, their decomposition shows that hypergraphs with equal size-labeled co-membership counts produce identical dynamics, which is the sufficiency of the probabilities $P_i(J,d)$; this consequence is not stated in Ref.~\cite{neuhauser}. Two elements are absent from that framework: the compression of $P_i(J,d)$ into inverse power moments, which exploits the polynomial dependence on $1/d$, and the necessity direction. The effective order is a property of a single pair of structure and rule. Rommens et al.~\cite{rommens} showed that, for binary states, the decomposition underlying it is fixed by the dynamical map in every exact representation; Lemma~\ref{lem:mobius} and Corollary~\ref{cor:single} are the stochastic, single-rule counterpart of that statement, whereas Theorem~\ref{thm:minimal} addresses a different question: which statistics of the group-interaction process are identifiable for a class of rules.

\paragraph{Information-theoretic reducibility.}
Lucas et al.~\cite{lucas} assess reducibility by comparing the density matrix of linear diffusion on a hypergraph with those of its truncations to lower orders, selecting the order through a complexity-penalized Kullback--Leibler divergence score. This is an approximate criterion for compressing a given hypergraph under a given dynamics, rather than an exact equivalence between different structures. Because the multiorder Laplacian is built from pairwise co-membership counts, resolved by hyperedge size, the criterion depends on the structure only through pairwise information, of the type contained in the probabilities $P_i(\{j\},d)$; joint co-membership of larger sets does not enter. In particular, the hypergraphs of Example~\ref{ex:octahedron} have identical multiorder Laplacians, hence identical density matrices and identical reducibility at every diffusion time, although they generate distinct quadratic voter dynamics. As expected for a criterion built on linear diffusion, it is by construction blind to structure that matters for nonlinear dynamics.

\paragraph{Representation, identifiability and description length.}
Rommens et al.~\cite{rommens} separate structural representation, functional representability, statistical identifiability and mechanistic adequacy, and prove that the interaction order of a node's update, together with its essential interaction sets for binary states, is a property of the dynamical map that no exact representation can alter. Their analysis concerns deterministic maps with a fixed rule; the present results are a stochastic, class-level counterpart in the second and third of these senses. For a single rule, Corollary~\ref{cor:single} identifies the minimal structural statistic; for a class of rules, Theorem~\ref{thm:minimal} shows that the identifiable object is the moment hierarchy $\Pi_q$ of the group-interaction process. The choice among equivalent encodings is then a question of description length rather than of correctness, and it is meaningful only relative to an explicit code. Under a code that repeats each label on its edges, a hyperarc with $|J|$ tails costs about $|J|$ times as much in the labeled-graph encoding, analogous to the factor $k-1$ that Rommens et al.\ obtain for clique projections of disjoint groups and to the comparison between hyperarcs and the projected graph in Ref.~\cite{burgio}, which can go either way depending on how the dependency sets overlap; at the algorithmic level, where no code is privileged, the encodings coincide up to a constant, and, as Rommens et al.\ conclude for bipartite and multilayer lifts, the choice turns on which entities a formalism takes as primitive.

\paragraph{Open questions.}
Comparing the description length of $\Pi_q$ with that of the underlying mechanism, within the minimum-description-length framework of Ref.~\cite{rommens}, remains open. So does the extension beyond binary states: for continuous variables the interaction order remains a property of the map~\cite{rommens}, but resolving it into lower-order terms requires a reference point~\cite{burgio}, and an analogue of Theorem~\ref{thm:minimal} would be needed to identify the corresponding statistics of the group-interaction process. Extensions to multi-state agents and state-dependent group selection raise the same issue, as does correlated group selection, for which the analogue of $\Pi_q$ would include cross-agent moments.

Group interactions are the phenomenon; graphs and hypergraphs are instruments for describing it. For a given class of dynamics, the results above specify which aspects of the interaction process an instrument must retain and which it may discard without loss, while making clear that what a dynamics cannot see is not thereby absent from the phenomenon. Conversely, dependencies among three or more agents in the node states do not by themselves reveal a group mechanism, since pairwise rules also produce them~\cite{villegas,perezreche}.

\section*{Acknowledgments}
The author acknowledges support from grant PID2024-158120NB-C22, funded by MICIU/\allowbreak AEI/\allowbreak 10.13039/\allowbreak 501100011033. The author thanks Lucas Lacasa for reading early drafts of this work and for fruitful comments.

\paragraph*{AI use.} The author used an AI assistant (Claude Opus~5.5, Anthropic) throughout the preparation of this work. The research question, the framework and the main results originate with the author, who also identified the relevant connections with the existing literature and directed the positioning of the paper. The assistant was used to check or improve derivations and examples, including verification in exact arithmetic; to search and summarize literature; to draft and revise text and, under the author's direction, to prepare the figure; and to provide critical reviews of successive drafts. All content was reviewed, corrected where necessary, and approved by the author, who takes full responsibility for the paper.

\end{document}